\documentclass[reqno,centertags,11pt]{amsart}
\usepackage{amsmath,amsthm,amscd,amssymb}
\usepackage{latexsym}
\usepackage{pstricks}
\usepackage{pst-node}

\usepackage {verbatim}

\newcommand{\N}{{\mathbb{N}}}

\newcommand{\R}{{\mathbb{R}}}

\newcommand{\C}{{\mathbb{C}}}

\newcommand{\no}{\nonumber}

\newcommand{\f}{\frac}

\newcommand{\ol}{\overline}

\newcommand{\wti}{\widetilde  }

\newcommand{\hatt}{\widehat}
\newcommand{\beq}{\begin{equation}}
\newcommand{\eeq}{\end{equation}}
\newcommand{\bdm}{\begin{displaymath}}
\newcommand{\edm}{\end{displaymath}}
\newcommand{\ba}{\begin{align}}
\newcommand{\ea}{\end{align}}
\newcommand{\bpf}{\begin{proof}}
\newcommand{\epf}{\end{proof}}
\newcommand{\eps}{\epsilon}

\newcommand{\la}{\langle}
\newcommand{\ra}{\rangle}

\newcommand{\supp}{\mathrm{supp}\, }               
\newcommand{\dist}{\mathrm{dist}}               

\newcommand{\e}{\mathrm{e}}

\newcommand{\veps}{\varepsilon}

\newcommand{\re}{\mathrm{Re}}
\newcommand{\im}{\mathrm{Im}}

\allowdisplaybreaks

\newcommand{\calC}{\mathcal{C}}

\newcommand{\calQ}{\mathcal{Q}}
\newcommand{\calR}{\mathcal{R}}

\newtheorem{thm}{Theorem}
\newtheorem{prop}[thm]{Proposition}
\newtheorem{lemma}[thm]{Lemma}
\newtheorem{cor}[thm]{Corollary}

\theoremstyle{definition}

\newtheorem{remark}[thm]{Remark}
\newtheorem{remarks}[thm]{Remarks}

\newcounter{theoremi}[thm]

\newcommand{\itemthm}{\refstepcounter{theoremi} {\rm(\roman{theoremi})}{~}}

\numberwithin{thm}{section}
\numberwithin{equation}{section}

\begin{document}

\title[Decay and smoothness of Dispersion Managed Solitons]{%
    Decay estimates and smoothness for solutions of the dispersion managed
    non-linear Schr\"odinger equation}
\author[D.~Hundertmark and Y.-R.~Lee]{Dirk Hundertmark$^1$ and Young-Ran~Lee$^2$}
\address{School of Mathematics, Watson Building, University of Birmingham, Edgbaston, Birmingham, B15 2TT, UK,
    on leave from Department of Mathematics, Altgeld Hall,
    University of Illinois at Urbana-Champaign, 1409 W.~Green Street, Urbana, IL 61801.}%
\email{dirk@math.uiuc.edu}%
\address{Department of Mathematical Sciences
KAIST (Korean Advanced Institute of Science and Technology),
335 Gwahangno, Yuseong-gu, Daejeon, 305-701, Republic of Korea.}%
\email{youngranlee@kaist.ac.kr}

\thanks{ $^1$ supported in part by NSF grant DMS--0400940}
\thanks{ $^2$ supported in part by the Korean Science and Engineering Foundation(KOSEF)
grant funded by the Korean government(MOST) (No.R01-2007-000-11307-0).}
\thanks{\copyright 2008 by the authors. Faithful reproduction of this article,
        in its entirety, by any means is permitted for non-commercial purposes}
\thanks{April 2008, version 15, file dms-15.tex}

\begin{abstract}
We study the decay and smoothness of solutions of the dispersion
managed non-linear Schr\"odinger equation in the case of
zero residual dispersion. Using new $x$-space versions of bilinear
Strichartz estimates, we show that the solutions are not only
smooth, but also fast decaying.
\end{abstract}

\maketitle

\section{Introduction}\label{sec1}
The parametrically excited one-dimensional non-linear Schr\"odinger
equation (NLS) with periodically varying dispersion
coefficient
 \beq\label{eq:periodic-NLS}
  i u_t + d(t) u_{xx} + c|u|^2 u = 0
 \eeq
arises naturally as an envelope equation for electromagnetic wave
propagation in optical wave\-guides used in fiber-optics
communication systems where the dispersion is varied periodically
along an optical fiber; it describes the amplitude of a signal
transmitted via amplitude modulation of a carrier wave through a
fiber-optical cable, see, e.g., \cite{Agrawal95,SulemSulem,Tetal03}.
In \eqref{eq:periodic-NLS} $t$ corresponds to
the distance along the fiber, $x$ denotes the (retarded) time,
$u_t=\partial_t u= \tfrac{\partial}{\partial t} u$, $u_{xx}=
\partial_x^2 u= \tfrac{\partial^2}{(\partial x)^2} u$, $c$ a constant determining
the strength of the non-linearity which, for convenience, we put equal to one
in the following, and $d(t)$ the dispersion along the waveguide, which,
for practical purposes, one can assume to be piecewise constant.
The balance between dispersion and non-linearity is the key factor which
determines the existence of stable soliton like pulses.

With fast data transfer through fiber-optic cables over long
intercontinental distances in mind, one would like to use stable
pulses, i.e., solitons, which do not change shape when traveling
through the cable. The NLS does support solitons, but those depend
on a delicate balance between dispersion and non-linearity. Also
these solitary pulses then strongly interact with each other via the
non-linear effects, which limits the bandwidth of the waveguide
since each pulse must, therefore, be well separated from the next.
Even worse, when multiplexing, that is, using multiple carrier waves
with different frequencies, blue, green, and red, say, to create
several channels in the waveguide which can be used simultaneously
to increase the bit rate through the fiber, the pulses in each
channel will travel with different group velocities determined by
the carrier frequency and the dispersion relation in the optical
cable. These pulses will overtake the ones on the `slower' channel,
hence pulses from different channels are bound to strongly interact.

One possibility to limit these negative effects of the non-linearity
is to stay in the linear regime, with vanishing non-linearity, or at
least the quasi-linear regime, where the non-linearity is small and
hence the pulses do not interact, at least not much, with each
other. On the other hand, there are no stable pulses in these
regimes of small non-linearity where the dispersion dominates. All
pulses broaden due to the dispersion, which again severely limits
the bandwidth of long optical waveguides.

The technique of dispersion management was invented to overcome the
difficulty that there are no stable pulses in the linear regime. The
idea, building on the fact that optical fibers can be engineered to
have positive and negative dispersion, see \cite{CLF79}, is to use
alternating sections of constant but opposite, or nearly opposite,
dispersion. This introduces a rapidly varying dispersion $d(t)$ along
the fiber, which, if the dispersion exactly cancel each other, leads
to pulses changing periodically along the fiber. This idea had been introduced
in 1980 in \cite{LKC80}. It has turned out to be enormously fruitful,
see for example, \cite{AB98, GT96a, GT96b, KH97, Kurtzge93, LK98, MM99, MMGNMG-NV99}
and the references therein, even if one takes small
non-linear effects into account and allows for a small residual
dispersion along the fiber, the residual dispersion together with
the non-linearity can balance each other, allowing the existence of
stable soliton-like pulses. Record breaking transmission rates of
more than 1 Tbits/s over an 18,000 kilometer optical fiber had been
achieved using this technology \cite{MGLWXK03} and the technique of
dispersion management is now widely used commercially.

Due to the enormous practical implications, there has been a huge
literature concerning the numerical and phenomenological
explanations and the theoretical, but most often non-rigorous,
understanding of the stabilizing effects of dispersion management
techniques, mainly in the regime of strong dispersion management. In
this regime neither the non-linearity nor the residual dispersion
need to be small, but they are small relative to the local
dispersion given by
 \beq\label{eq:local-dispersion}
    d(t) = \frac{1}{\veps} d_0(t) +d_{\text{av}} .
 \eeq
Here $d_0(t)$ is the mean zero part, $d_\text{av}$ the average dispersion
over one period, and $\veps$ a usually small parameter. The envelope equation
valid in this regime was derived by Gabitov and Turitsyn in 1996, \cite{GT96a,GT96b}.
It is given by a non-linear Schr\"odinger equation which, after rescaling $t$ to $t/\veps$,
takes the form
 \beq\label{eq:GT}
  i u_t + d_0(t) u_{xx} + \veps(d_{\text{av}}u_{xx}+ |u|^2 u ) =0 .
 \eeq
Note that the average dispersion and non-linearity is small compared to the
local dispersion, which is a characteristic feature of the strong dispersion
management regime.

Since the full equation \eqref{eq:GT} is very hard to study, one
makes one approximation to the full equation: Assume that the mean
zero part of dispersion along the fiber is $-1$ on the interval
$[-1,0]$ and $+1$ on $[0,1]$. Then separating the free motion given
by the solution of $-iu_t +d_0(t)u_{xx}=0$ and averaging
\eqref{eq:GT} over the period, see \cite{AB98,GT96a}, yields
 \beq\label{eq:DMS-time}
  i v_t +\veps d_{\text{av}} v_{xx} + \veps Q(v,v,v) = 0
 \eeq
for the ``averaged'' solution $v$, where
  \beq\label{eq:Q3}
   Q(v_1,v_2,v_3) :=  \int_0^1 T_r^{-1}\bigl[ {T_rv_1} \ol{T_rv_2} T_rv_3 \bigr] dr
  \eeq
and $T_r= e^{ir\partial_x^2}$ is the solution operator of the free
Schr\"odinger equation and, by symmetry, we restrict the integration
in the $r$-variable to $[0,1]$. In some sense, $v$ is a
slowly varying variable along the optical waveguide and the varying dispersion, is
interpreted, in the spirit of Kapitza's treatment of the unstable
pendulum which is stabilized by fast small oscillations of the
pivot, see \cite{LandauLifshitz}, as a fast background oscillation,
justifying formally the above averaging procedure.

The Gabitov--Turitzin model \eqref{eq:DMS-time} for the dispersion
managed optical waveguide is well-supported by numerical studies, see,
for example, \cite{AB98} and \cite{TDNMSF99}, and theoretical
arguments, see, for example, \cite{Lushnikov01, Lushnikov04}. In
addition, this averaging procedure was rigorously justified in
\cite{ZGJT01} where it is shown that in the regime of strong
dispersion management, $\veps\ll 1$, on long scales $0\le t\le
C\veps^{-1}$ the solution of the full equation \eqref{eq:GT} stays
$\veps$-close to a solution of \eqref{eq:DMS-time} with the same
initial data, showing that it is indeed an infinite dimensional
analogue of Kapitza's effect. Moreover, the averaged equation
\eqref{eq:DMS-time} supports stable solitary solutions in certain
regimes of the parameters. These solitary solutions give the average
profile of the breather--like pulses in \eqref{eq:periodic-NLS}.
Making the ansatz $v(t,x)= e^{i\veps\omega t} f(x)$  in
\eqref{eq:DMS-time} yields the time independent equation
 \beq\label{eq:DMS}
  -\omega f = -d_{\text{av}}f_{xx} -Q(f,f,f)
 \eeq
describing stationary soliton-like solutions, the so-called dispersion managed
solitons. Equation \eqref{eq:DMS} is the Euler-Lagrange equation for
the averaged Hamiltonian
 \beq\label{eq:averaged-energy}
    H(f) = \f{d_{\text{av}}}{2} \int_{\R}|f'|^2dx - \f{1}{4} \calQ(f,f,f,f),
 \eeq
where we set
 \beq\label{eq:Q4}
  \calQ(f_1,f_2,f_3,f_4)= \int_0^1\int_{\R} \ol{T_rf_1(x)} T_rf_2(x)
  \ol{T_rf_3(x)} T_rf_4(x)\, dx dr.
 \eeq
Again $T_r$ is the free Schr\"odinger evolution.

The very large literature of numerical and phenomenological
explanations and the theoretical understanding of the stabilizing
effects of dispersion management techniques in the strong dispersion
management regime is mainly based on the averaged equation
\eqref{eq:DMS}. Despite this enormous interest in dispersion managed
solitons, there are few rigorous results available. Note that
both $Q(f,f,f)$ and $\calQ(f,f,f,f)$ are nonlinear and, in
addition, highly non-local functions of $f$. This presents a
unique challenge in the study of \eqref{eq:DMS}. Existence of
solutions of \eqref{eq:DMS} had first been rigorously established in
\cite{ZGJT01} for positive residual dispersion $d_{\text{av}}>0$.
Instead of showing existence of solutions of \eqref{eq:DMS}
directly, the existence of minimizers of the constraint minimization
problem
 \beq\label{eq:minimization1}
  \ol{P}_{\lambda,d_\text{av}}= \inf\Big\{ H(f):\, f\in H^1(\R), \|f\|_2^2= \lambda \Big\}
 \eeq
was proved. Simple Gaussian testfunctions show $\ol{P}_{\lambda,d_\text{av}}<0$. Since
\eqref{eq:DMS} is the Euler-Lagrange equation for constraint
minimization problem \eqref{eq:minimization1}, any minimizer of
\eqref{eq:minimization1} is a weak solution of \eqref{eq:DMS} with
some $\omega>0$. Of course, minimizing sequences for the
minimization problem \eqref{eq:minimization1} can very easily
converge weakly to zero, since the functional
\eqref{eq:averaged-energy} is invariant under shifts of $f$. This
non-compactness was overcome using Lions' concentration compactness
principle \cite{Lions84}.

In the case of positive $d_{\text{av}}$, every weak solution $f\in
H^1(\R)$ of \eqref{eq:DMS} with $\omega >0$ is automatically $\calC^\infty$;
recall that $\omega >0$ for any minimizer of \eqref{eq:minimization1}. The
smoothness follows from a simple bootstrapping argument. Using that $f\mapsto
Q(f,f,f)$ maps the Sobolev spaces $H^s(\R)$ into themselves and
$(\omega -d_{\text{av}}\partial_x^2)^{-1}$ maps $H^s(\R)$ into $H^{s+2}(\R)$,
as long as $\omega >0$, straightforward bootstrapping shows that
any solution $f\in H^1(\R)$ of \eqref{eq:DMS} with $\omega>0$ is in all the
$H^s(\R)$ Sobolev spaces for all $s\ge 1$, hence smooth by the Sobolev
embedding theorem.

The variational problem in the case of vanishing residual
dispersion, $d_{\text{av}}=0$ is much more subtle and complicated
due to an additional loss of compactness. Nevertheless, it is very
important physically, since certain physical effects which
destabilize pulse propagation in optical fibers are minimal for
$d_{\text{av}}$ near or equal to zero \cite{SLGKPST02,THFKS99}. In this case
the constraint minimizing problem is given by
 \beq\label{eq:minimization-critical}
  \ol{P}_{\lambda} = \inf\Big\{ -\frac{1}{4}\calQ(f,f,f,f):\, f\in L^2(\R), \| f\|_2^2= \lambda\Big\}.
 \eeq
Using the Strichartz inequality, it was shown in \cite{ZGJT01} that
even in this case  $\ol{P}_\lambda>-\infty$, see also Lemma \ref{lem:0}
below. Now the minimizing sequence is only bounded in $L^2$ and, since
the functional in \eqref{eq:minimization-critical} is invariant under shift
of $f$ in real space and in Fourier space, the traditional
a-priori bounds from the calculus of variations are not available.
The existence of a minimizer for the variational problem \eqref{eq:minimization-critical}
was shown by Markus Kunze \cite{Kunze04}, using
the concentration compactness principle in tandem; first in Fourier
and then in $x$-space. This minimizer yields a solution for dispersion management
equation \eqref{eq:DMS} for vanishing average dispersion, $d_\text{av}=0$.
Unfortunately, the bootstrapping argument
which shows smoothness of solutions of \eqref{eq:DMS} for
$d_{\text{av}}\!>\!0$ now fails when $d_{\text{av}}\!=\!0$ since there is a loss of
the second order derivatives. The minimizer is
now only in $L^2(\R)$ and the nonlinearity $Q$ is not smoothness improving,
so Kunze's method does not give much more a-priori
information on the minimizer besides being square integrable and
bounded. Shortly afterwords, Milena Stanislavova showed that Kunze's
minimizer is smooth. Her approach employed the use of Bourgain
spaces \cite{Bourgain1, Bourgain2} and Tao's bilinear estimates
\cite{Tao}.

To the best of our knowledge these results are the only known
rigorous results concerning solutions of \eqref{eq:DMS}. For
example, nothing is rigorously known so far on the decay properties
of dispersion managed solitons. This is a tantalizing situation:
since the tails of dispersion managed solitons are responsible for
the interactions of pulses launched into the optical fiber, the
tails essentially limit the bit rate capacity of optical waveguides.
Thus finding the asymptotic behavior of dispersion managed solitons
is an important fundamental and practical problem which has
attracted a lot of attention in numerical and phenomenological
studies. Lushnikov \cite{Lushnikov04} gave convincing but
non-rigorous arguments that for any solution $f$ of \eqref{eq:DMS},
 \beq
  f(x)\sim A\cos(a_0x^2 + a_1 x + a_2) \e^{-b|x|} \quad \text{as }
  x\to\infty
 \eeq
for some suitable choice of real constants $a_j $ and $b>0$, see
also \cite{Lushnikov01}.

In this paper we derive \emph{the first rigorous decay bounds} on
dispersion managed solitons. Although our approach is, so far, not
able to give exponential decay of dispersion managed solitons
conjectured by Lushnikov, it shows that any solution of
\eqref{eq:DMS} in the case of vanishing residual dispersion,
$d_{\text{av}}=0$, is super--polynomially decaying.

\begin{thm}\label{thm:main}
Let $\omega>0$. Any weak solution $f\in L^2(\R)$ of $\omega f=
Q(f,f,f)$ is a Schwarz function. That is, $f$ is arbitrary often
differentiable and $f$ and all its derivatives $f^{(n)}$ decay
faster than polynomially at infinity,
 \bdm
 \sup_{x} |x|^m |f^{(n)}(x)| <\infty \quad \text{ for all }
 m,n\in\N_0.
 \edm
\end{thm}
\begin{remarks}\itemthm\label{rem:weak-solution} By a weak solution, we mean
 a function $f\in L^2(\R)$ such that
 \beq
  \la g,f\ra = \la g, Q(f,f,f)\ra = \calQ(g,f,f,f)
 \eeq
 for all $g\in L^2(\R)$. Here $\la g, f\ra= \int_\R \ol{g(x)}f(x)\, dx$
 is the usual scalar product on $L^2(\R)$. Note that our scalar product
 is sesquilinear in the first component and linear in the second. Also,
 due to the Strichartz inequality, the functional $\calQ(f_1,f_2,f_3,f_4)$
 is well-defined as soon as $f_j\in L^2(\R)$ for all $j=1,2,3,4$,
 see Lemma \ref{lem:0}. In turn, this
 means that $Q(f_1,f_2,f_3)$ is an $L^2$ function for all
 $f_1,f_2,f_3\in L^2(\R)$ and the notion of a weak solution of
 \eqref{eq:DMS} by testing with $L^2$ functions, if $d_{\text{av}}=0$,
 respectively, $H^1(\R)$ functions if $d_{\text{av}}>0$, makes
 sense. \\[0.2em]
\itemthm Since $\calQ(f,f,f,f)=0$ implies $f\equiv 0$ by the unicity of $T_t$, any
nontrivial weak solution of $\omega f = Q(f,f,f)$ automatically has
$\omega = \omega_f=\calQ(f,f,f,f)/\la f,f\ra>0$. \\[0.2em]
\itemthm One can give a more precise estimate on the super-polynomial
decay rate of $f$, see Remark \ref{rem:f-decay} and
Corollary \ref{cor:arbitrary-polyn-decay}. This misses the
conjectured exponential decay rate, however.\\[0.2em]
\itemthm Theorem \ref{thm:main} significantly
strengthens Stanislavova's result on smoothness of dispersion
managed solitons in \cite{Stanislavova}. In addition, our proof is
technically much simpler than Stanislavova's.
\end{remarks}

We will deduce the regularity property of dispersion managed
solitons given in Theorem \ref{thm:main} from a suitable decay
estimate on the tails of the solutions $f$ and its Fourier transform
$\hatt{f}$. For this we need some more notations. For $f\in
L^2(\R)$, let
 \begin{align}
  \alpha(s)
 &:=
  \big( \int_{|x|\ge s} |f(x)|^2 \, dx\big)^{1/2}\\
  \beta(s)
 &:=
  \big( \int_{|k|\ge s} |\hatt{f}(k)|^2 \, dk\big)^{1/2}
 \end{align}
be the $L^2$-norm of its tail, respectively the tail of its Fourier
transform $\hatt{f}$. For a general function $f\in L^2(\R)$, the
only thing one can say a-priori about $\alpha$ and $\beta$
is that they both decay to zero as $s\to\infty$. In general, this
decay can be arbitrarily slow. For weak solutions of the dispersion
management equation more is true.
\begin{prop}[Super-algebraic decay of the tails]\label{prop:decay}
Any weak solution $f\in L^2(\R)$ of $\omega f=Q(f,f,f)$ obeys the
a-priori estimates
 \begin{align}
  \alpha(s)&\le C_\gamma s^{-\gamma} \\
  \beta(s)&\le C_\gamma s^{-\gamma}
 \end{align}
for all $s>0$, all $\gamma>0$, and some finite constant $C_\gamma$.
\end{prop}
\begin{remarks}\itemthm In fact, a slightly stronger result holds,
see Corollary \ref{cor:arbitrary-polyn-decay}.\\[0.2em]
\itemthm\label{rem:f-decay} We get a decay estimate for $f$  similar to
the one for $\alpha$, since
 \bdm
 \begin{split}
  |f(s)|^2 +|f(-s)|^2 &= 2\int_{|x|>s} \re{f'(x)f(x)}\, dx
    \le 2\int_{|x|>s} |f'(x)f(x)|\, dx
    \le 2\|f'\|_2^2 \alpha(s) .
  \end{split}
 \edm
\end{remarks}
An immediate corollary of Proposition \ref{prop:decay} is that for any weak solution
of $\omega f=Q(f,f,f) $ both $f$ and $\hat{f}$ are in the Sobolev
spaces $H^s(\R)$ for arbitrary $s\ge 0$, in particular, both $f$ and
$\hat{f}$ are infinitely often differentiable.
\begin{prop}\label{prop:smooth}
Any weak solution $f\in L^2(\R)$  of $\omega f= Q(f,f,f)$ is  in the
Sobolev space $H^s(\R)$ for any $s\ge 0$. The same holds for
$\hat{f}$. In particular, $f$ and $\hat{f}$ are in
$\calC^\infty(\R)$.
\end{prop}
In turn, Theorem \ref{thm:main} is a direct consequence of
Proposition \ref{prop:smooth}, see Lemma \ref{lem:Schwartz}.
In the next section we establish our main technical tools, multi-linear
refinements of Strichartz estimates both in Fourier space and in $x$-space,
Corollary \ref{cor:4linear}, and the quasi-locality of the non-local
functional $\calQ$, Lemma \ref{lem:quasi-locality}. In Section
\ref{sec:proof-main} we use the above results to prove
self-consistency bounds on the tail-distributions, Lemma
\ref{lem:tail-estimate}. These self-consistency bounds are similar
in spirit to sub-harmonicity bounds and are the main tool in our
proof of the super-algebraic decay of dispersion managed solitons given in
Corollary \ref{cor:arbitrary-polyn-decay}, which is a refinement
of Proposition \ref{prop:decay}.

\section{Multi-linear estimates} \label{sec:multilinear}
We want to study the smoothness and decay properties of general
solutions of the (averaged) dispersion management equation
\eqref{eq:DMS} in the case of vanishing average dispersion,
$d_{\text{av}}=0$. That is, we assume that $f\in L^2(\R)$ is a weak
solution of
 \beq\label{eq:DMS-critical}
  \omega f = Q(f,f,f)
 \eeq
with $Q(f,f,f)$ given by \eqref{eq:Q3}. As mentioned in Remark
\ref{rem:weak-solution}, the right hand side of
\eqref{eq:DMS-critical} is an $L^2(\R)$ function for any $f\in
L^2(\R)$, thus the notion of a weak solution of
\eqref{eq:DMS-critical} makes sense; $f\in L^2(\R)$ is a weak
solution of \eqref{eq:DMS-critical} if
 \beq\label{eq:DMS-weak}
  \omega \la g,f\ra = \la g, Q(f,f,f) \ra = \calQ(g,f,f,f)
 \eeq
for all $g\in L^2(\R)$. The second equality in \eqref{eq:DMS-weak}
follows from the unicity of $T_t= e^{it\partial_x^2}$. Since
$\calQ(f,f,f,f)>0$ for $f\not\equiv 0$, $\omega>0$.

The ground-state dispersion managed soliton is a solution of the
minimization problem \eqref{eq:minimization-critical} or,
equivalently, the maximization problem
 \beq\label{eq:maximization}
  P_\lambda =\sup \Big\{\calQ(f,f,f,f):\, f\in L^2(\R), \| f\|_2^2=
  \lambda\Big\} .
 \eeq
By scaling, $\wti{f}= f/\sqrt{\lambda}$, one sees $P_\lambda= P_1
\lambda^2$. Thus if $f$ is a ground-state dispersion managed
solitons then, testing \eqref{eq:DMS-critical} with $f$, one sees
that $f$ solves \eqref{eq:DMS-critical} with $\omega$ given by
 \beq\label{eq:omega-ground-state}
  \omega = P_1 \lambda = P_1 \|f\|_2^2 .
 \eeq

We give several preparatory lemmas in this section.
\begin{lemma}\label{lem:0}
The two-sided bound $ 1.05 (2\pi)^{-1/2} \le P_1 \le 12^{-1/4} $
holds. In particular, for any functions $f_j\in L^2(\R)$, j=1,2,3,4,
 \beq\label{eq:Q}
  |\calQ(f_1,f_2,f_3,f_4)| \le P_1 \prod_{j=1}^4 \| f_j\|
  \le
  12^{-1/4} \prod_{j=1}^4 \| f_j\| .
  \eeq
\end{lemma}
\begin{proof} Using the triangle and generalized H\"older inequalities,
  \begin{align*}\label{eq:Hoelder}
  | \calQ(f_1,f_2,f_3,f_4) |
  & \leq \int_0^1 \int _{\R} \prod _{j=1}^4 |T_tf_j| \, dx dt\\
 & \leq \prod _{j=1}^4 \left(\int_0^1 \int _{\R}
 |T_t f_j|^4 dx dt \right)^{1/4}\,
  = \prod _{j=1}^4 \left( \calQ (f_j, f_j, f_j, f_j )\right)^{1/4} .
  \end{align*}
Given $f$ let $\wti{f}=f/\|f\|_2$, by definition of $P_1$,
 \bdm
   \calQ(f, f, f, f )
   =
   \calQ(\wti{f}, \wti{f}, \wti{f}, \wti{f} ) \|f\|_2^4
   \le P_1 \|f\|_2^4
 \edm
which gives the first inequality in \eqref{eq:Q}. The second
inequality in \eqref{eq:Q} follows once the upper bound on $P_1$ is
proven. For this we use the one-dimensional Strichartz inequality,
 \beq\label{eq:Strichartz}
  \int_{\R}\int_{\R} |T_tf(x)|^6 \, dx dt \le S_1^6 \|f\|^6_{L^2(\R)},
 \eeq
which holds due to the dispersive properties of the free
Schr\"odinger equation, \cite{GV85,Strichartz,SulemSulem}.  The sharp
constant in \eqref{eq:Strichartz} is known, $S_1= 12^{-1/12}$, one
even knows $S_2$ in two space dimensions, see \cite{Foschi,HZ06},
but, so far, not in any other space dimension $d\ge3$. Let $\|f\|_2=1$.
Using the Cauchy-Schwarz inequality, one gets
 \bdm
  \calQ(f,f,f,f)= \int_0^1\int_{\R} |T_tf|^{3+1} dxdt
  \le
  \bigl(\int_0^1\int_{\R} |T_tf|^6 dxdt\bigr)^{1/2}
  \bigl(\int_0^1\int_{\R} |T_tf|^2 dxdt \bigr)^{1/2} .
 \edm
The first factor is bounded with the help of \eqref{eq:Strichartz}
by extending the integral in $t$ to all of $\R$. The second factor
is bounded by doing the $x$-integration first, using that $T_t$ is a
unitary operator on $L^2(\R)$. Thus
 \bdm
  \calQ(f,f,f,f)
 \le
  S_1^3 \quad \text{for all } \|f\|_2=1 .
 \edm
Hence $P_1\le S_1^3= 12^{-1/4}$, using the sharp value for the
Strichartz constant. This proves the upper bound on $P_1$ and thus
the second inequality in \eqref{eq:Q}.

For the lower bound on $P_1$, we use a chirped Gaussian
test-function similar to \cite{Kunze04, ZGJT01}. If the initial
condition $f$ is given by
 \beq
  f(x)= A_0 e^{-x^2/\sigma_0} \quad \text{ with }\re(\sigma_0)>0
 \eeq
then with $\sigma(t)= \sigma_0+4it$ and $A(t)=
A_0\sqrt{\sigma_0}/\sqrt{\sigma(t)}$, its free time-evolution is
given by
 \beq
  T_tf(x) = A(t) e^{-x^2/\sigma(t)},
 \eeq
see, e.g., \cite{ZGJT01}. Thus
 \beq
  \int_0^1\int_{\R} |T_tf|^4\, dx dt = \sqrt{\frac{\pi}{4}}
  |A_0|^4|\sigma_0|^2 \int_0^1 \frac{1}{|\sigma(t)|}\, dt .
 \eeq
Choosing $|A_0|^2= \sqrt{2\re(\sigma_0)/(|\sigma_0|^2)\pi}$ yields
the normalization  $\|f\|_2=1$ and hence
 \beq
  P_1
 \ge
  \frac{\re(\sigma_0)}{\sqrt{\pi}}
  \int_0^1\frac{1}{\sqrt{\re(\sigma_0)^2+(\im(\sigma_0)+4t)^2}}\, dt
  .
 \eeq
The best choice for $\im(\sigma_0)$ is $\im(\sigma_0)=-2$ and with
$\delta= 2/\re(\sigma_0)$ we arrive at
 \beq
  P_1
 \ge
  \frac{1}{\sqrt{2\pi}} \sup_{\delta>0}
  \frac{1}{\sqrt{\delta}}\int_0^{\delta}\frac{1}{\sqrt{1+s^2}}\, ds
  >
  \frac{1.05}{\sqrt{2\pi}} ,
 \eeq
which, noticing that the supremum is attained at approximately $\delta= 3.32$,
gives the claimed lower bound on $P_1$.
\end{proof}

Besides some estimates on $\calQ$, we also need, for technical
reasons, bounds on the slightly modified functional
 \beq\label{eq:calR}
 \calR(f_1,f_2,f_3,f_4) :=
 \int_0^1 \int_{\R} \ol{T_tf_1(x)}T_tf_2(x)\ol{T_tf_3(x)}T_tf_4(x) \, tdxdt ,
 \eeq
where the measure $dx dt$ on $\R\times[0,1]$ is changed to $tdxdt$.
\begin{lemma}\label{lem:1} For any functions $ f_j\in L^2(\R)$, j=1,2,3,4,
 \beq\label{eq:R}
  |\calR(f_1,f_2,f_3,f_4)| \le \frac{12^{-1/4}}{ \sqrt{3}} \prod_{j=1}^4 \| f_j \| .
 \eeq
\end{lemma}
\bpf Again, as in the proof of Lemma \ref{lem:0}, using the triangle
and generalized H\"older inequalities, one sees
 \bdm
  |\calR(f_1,f_2,f_3,f_4) |
  \le \prod _{j=1}^4 \calR\left(f_j, f_j, f_j, f_j \right)^{1/4} .
 \edm
So it is enough to prove \eqref{eq:R} in the case $f_j=f$ for all
$j=1,2,3,4$. Using the Cauchy-Schwarz inequality,
 \bdm
  \calR(f,f,f,f)= \int_0^1\int_{\R} |T_tf|^{3+1} t\, dxdt
  \le
  \Big(\int_0^1\int_{\R} |T_tf|^6\, dxdt \Big)^{1/2}
  \Big(\int_0^1\int_{\R} |T_tf|^2 t^2\, dxdt \Big)^{1/2} .
 \edm
Again, the first factor is bounded by extending the $t$-integration
to all of $\R$ and then using the one-dimensional Strichartz
inequality \eqref{eq:Strichartz} and the second doing the
$x$-integration first, using the unicity of $T_t$. Thus
 \bdm
  \calR(f,f,f,f)
 \le
  \frac{S_1^3}{\sqrt{3}}  \| f\|_{L^2(\R)}^4.
 \edm
Since $S_1^3=12^{-1/4}$, this proves \eqref{eq:R}. \epf

The following estimates are our main tools to prove the regularity
properties of dispersion managed solitons. The results below have
natural generalizations to arbitrary dimension. We need only their
one-dimensional versions.  Recall that $T_t= e^{it\partial_x^2}$ is
the solution operator for the free Schr\"odinger equation in
dimension one, that is
 \begin{align}
   T_t f(x)
  &=
   \frac{1}{\sqrt{4\pi it}} \int_{\R} e^{i\frac{|x-y|^2}{4t}} f(y)\, dy
   \label{eq:freetimeevolution1} \\
  &=
   \frac{1}{\sqrt{2\pi}} \int_{\R} e^{ix\eta} e^{-it\eta^2} \hatt{f}(\eta)\,
   d\eta .
   \label{eq:freetimeevolution2}
 \end{align}
 Here $\hatt{f}$ is the Fourier transform of $f$, given by
  \beq\label{eq:Fourier}
   \hatt{f}(\eta)
   =
   \frac{1}{\sqrt{2\pi}} \int_{\R} e^{-ix\eta} f(x)\, dx ,
  \eeq
for $f\in L^1(\R)\cap L^2(\R)$ and extended to a unitary operator to
all of $L^2(\R)$. The inverse Fourier transform is given by
$\check{f}$,
 \beq
 \check{ f} (x):= \f{1}{\sqrt{2\pi}}\int_{\R} e^{ix\eta}f(\eta)d\eta.
 \eeq
The following bilinear estimate for initial conditions $f_1$ and
$f_2$ whose Fourier transforms have separated supports is well
known.

\begin{lemma}[Fourier space bilinear estimate]\label{lem:bilinear-Fourier}
  If the initial conditions $f_1,f_2\in L^2(\R)$ have separated supports
  in Fourier space, $\dist (\supp \hatt{f_1}, \supp \hatt{f_2})>0$, then
  \beq\label{eq:bilinear-Fourier-space}
  \| T_tf_1 T_tf_2\|_{L^2(\R\times\R, dtdx)} \le
  \frac{1}{\sqrt{2\, \dist (\supp \hatt{f_1}, \supp \hatt{f_2})}}
  \| f_1\|_{L^2(\R)} \| f_2\|_{L^2(\R)} .
  \eeq
\end{lemma}
The above bound is one of the key ingredients to prove the
Fourier-space part of Theorem \ref{thm:main}. For the $x$-space
bounds, we need an $x$-space version of the above bilinear estimate.
For this the following observation is helpful.

\begin{lemma}[Duality]\label{lem:duality-bilinear}
Let $f_1,f_2\in L^2(\R)$ with $\check{f}_1$ and $\check{f}_2$ the
corresponding inverse Fourier transforms. Then
 \beq\label{eq:duality-bilinear}
  \| T_tf_1 T_t f_2\|_{L^2(\R\times \R, |t|^{-1} dt dx)}
  =
  \sqrt{2} \| T_t\check{f_1} T_t \check{f_2}\|_{L^2(\R\times \R,dtdx)}.
 \eeq
\end{lemma}
\begin{remark}
 Note that in the $L^2$-norm on the left hand side, the measure $dtdx$ is replaced by the measure
 $|t|^{-1}dt dx$, which is highly singular at $t=0$.
\end{remark}

Together with Lemma \ref{lem:bilinear-Fourier}, this duality result gives a real
space version of the bilinear estimates.

\begin{lemma}[$x$-space bilinear estimate]\label{lem:bilinear-real}
  If the initial conditions $f_1,f_2\in L^2(\R)$ have separated supports,
  $\dist (\supp f_1, \supp f_2)>0$, then
  \beq\label{eq:bilinear-real-space}
  \| T_tf_1 T_tf_2\|_{L^2(\R\times\R, |t|^{-1}dtdx)} \le
  \frac{1}{\sqrt{\dist (\supp f_1, \supp f_2)}} \| f_1\|_{L^2(\R)} \| f_2\|_{L^2(\R)}.
  \eeq
\end{lemma}

\begin{proof}[Proof of Lemma \ref{lem:bilinear-real}] Assuming Lemma
\ref{lem:bilinear-Fourier} and \ref{lem:duality-bilinear} for the
moment, the proof is straightforward. Since the Fourier transform of
$\check{f}_j$ is $f_j$ for  $j=1,2$, the assumption that $f_1$ and
$f_2$ have well separated supports, $\dist (\supp f_1, \supp
f_2)>0$, means that the supports of the Fourier transforms of
$\check{f}_1$ and  $\check{f}_2$ are well separated. Thus Lemma
\ref{lem:bilinear-Fourier} applies to the right hand side of
\eqref{eq:duality-bilinear} and hence
 \bdm
 \begin{split}
   \| T_tf_1 T_tf_2\|_{L^2(\R\times\R, |t|^{-1}dtdx)}
  &=
   \sqrt{2} \| T_t\check{f_1} T_t \check{f_2}\|_{L^2(\R\times \R, dt dx)} \\
  &\le
   \frac{1}{\sqrt{\dist (\supp f_1, \supp f_2)}} \| f_1\|_{L^2(\R)} \| f_2\|_{L^2(\R)}.
 \end{split}
 \edm
\end{proof}

It remains to prove the duality Lemma and the bilinear estimate in
Fourier space.

\begin{proof}[Proof of Lemma \ref{lem:duality-bilinear}]
Using the explicit form of the free time evolution
\eqref{eq:freetimeevolution1}, we see that
 \begin{align}
  &\int_{\R} \int_{\R} |T_tf_1T_tf_2|^2 \, |t|^{-1} dxdt \nonumber \\
  & =
  \f{1}{(4\pi)^{2}}
  \int_{\R} \int_{\R}
   \Big|
    \int_{\R^{2}} e^{ix(y_1+y_2)/(2t)} e^{-i(y_1^2+y_2^2)/(4t)}f_1(y_1)f_2(y_2)\, dy_1dy_2
   \Big|^2
  \, \f{dxdt}{|t|^{3}} \nonumber \\
  &=
  2 \int_{\R} \int_{\R}
   \Big|
    \f{1}{2\pi}\int_{\R^2} e^{iz(y_1+y_2)} e^{-i\tau (y_1^2+y_2^2)}f_1(y_1)f_2(y_2)\,
    dy_1dy_2
   \Big|^2 \, dz d\tau \label{eq:first-real-space}
 \end{align}
where we first made the change of variables $x= 2t z$, $dx= 2|t|dz$, and then
$t = 1/(4\tau)$ with $t^{-2}dt = 4 d\tau$. Let $\check{f_j}$ be the inverse
Fourier transform of $f$. Since
 \bdm
  T_\tau \check{f_j}(z) =
  \frac{1}{(2\pi)^{1/2}}
  \int_{\R} e^{iz y } e^{-i\tau y^2} f_j(y)\, dy
 \edm
one has
 \bdm
   T_\tau \check{f}_1(z) T_\tau \check{f}_2(z)
   =
   \f{1}{2\pi}\int_{\R^2} e^{iz(y_1+y_2)} e^{-i\tau (y_1^2+y_2^2)}f_1(y_1)f_2(y_2)\, dy_1dy_2
 \edm
and plugging this back into \eqref{eq:first-real-space} gives
 \bdm
  \int_{\R} \int_{\R} |T_tf_1T_tf_2|^2 \, |t|^{-1}dxdt
  =
  2\int_{\R}\int_{\R} |T_\tau \check{f_1}(z)T_\tau\check{f_2}(z)|^2\, dz d\tau
 \edm
which is \ref{eq:duality-bilinear}.
\end{proof}

\begin{proof}[Proof of Lemma \ref{lem:bilinear-Fourier}] This result is
known to the experts,  see, for example \cite{CKSTT01,OT98}. We
give a proof for the convenience of the reader. Using the Fourier
representation \eqref{eq:freetimeevolution2} of a solution of the
free Schr\"odinger equation,
 \bdm
  T_tf_1(x) T_tf_2(x) =
  \f{1}{2\pi} \int_{\R^{2}}
    e^{ix(k_1+k_2)} e^{-it(k_1^2+k_2^2)}\hatt{f}_1(k_1)\hatt{f}_2(k_2)\, dk_1dk_2 .
 \edm
In particular,
 \begin{multline*}
  \int_{\R} \int_{\R} |T_tf_1T_tf_2|^2 \, dxdt  = \\
  \f{1}{(2\pi)^{2}}
  \int_{\R} \int_{\R}
   \Big|
    \int_{\R^{2}}
      e^{ix(k_1+k_2)} e^{-it(k_1^2+k_2^2)}\hatt{f}_1(k_1)\hatt{f}_2(k_2)
    \, dk_1dk_2
   \Big|^2
  \, dxdt .
 \end{multline*}
Expanding the square, using $\delta(k)= \tfrac{1}{2\pi}\int_{\R}
e^{isk}\,ds$ as distributions, this leads to
\begin{multline}\label{eq:second-Fourier-space}
  \int_{\R} \int_{\R} |T_tf_1T_tf_2|^2\, dxdt = \\
  \int_{\R^{2}} \int_{\R^{2}}
   \delta(\eta_1+\eta_2- \zeta_1-\zeta_2)
   \delta(\eta_1^2+\eta_2^2 - \zeta_1^2-\zeta_2^2)
    \ol{\hatt{f}_1(\eta_1)} \ol{\hatt{f}_2(\eta_2)}
    \hatt{f}_1(\zeta_1)\hatt{f}_2(\zeta_2)
  \, d\eta_1d\eta_2d\zeta_1d\zeta_2 .
 \end{multline}

Now we make the change of variables $\xi_{1}= \eta_{1}+\eta_{2}$,
$\vartheta_{1}=\eta_{1}^2 + \eta_{2}^2$, and $\xi_{2}=
\zeta_{1}+\zeta_{2}$, $\vartheta_{2}=\zeta_{1}^2 + \zeta_{2}^2$. By the
inverse function theorem, the inverse of the Jacobian of the change of
variables $(\xi_1,\vartheta_1)\mapsto(\eta_1,\eta_2)$ is given by
$J^{-1}= \left(\begin{smallmatrix} \partial\xi_1/\partial\eta_1 & \partial\xi_1/\partial\eta_2\\
   \partial\vartheta_1/\partial\eta_1 & \partial\vartheta_1/\partial\eta_2\end{smallmatrix}\right)
   = \left(\begin{smallmatrix} 1 & 1 \\ 2\eta_1 & 2\eta_2 \end{smallmatrix}\right)$.
That is, $\det J^{-1}= 2(\eta_2-\eta_1)$ and hence
 \bdm
   d\eta_1d\eta_2 = |\det J| \, d\xi_1 d\vartheta_1 = \frac{d\xi_1 d\vartheta_1}{2|\eta_2-\eta_1|} .
 \edm
Thus, setting
$\hatt{f}_1\otimes\hatt{f}_2(\xi_1,\vartheta_1)=
\hatt{f}_1(\eta_1(\xi_1,\vartheta_1)) \hatt{f}_2(\eta_2(\xi_1,\vartheta_1))$
and similarly for $\hatt{f}_1\otimes\hatt{f}_2(\xi_2,\vartheta_2)$,
we can rewrite \eqref{eq:second-Fourier-space} as
 \begin{align*}
   & \int_{\R^{2}} \int_{\R^{2}}
   \delta(\eta_1+\eta_2- \zeta_1-\zeta_2)
   \delta(\eta_1^2+\eta_2^2 - \zeta_1^2-\zeta_2^2)
    \ol{\hatt{f}_1(\eta_1)} \ol{\hatt{f}_2(\eta_2)}
    \hatt{f}_1(\zeta_1)\hatt{f}_2(\zeta_2)
  \, d\eta_1d\eta_2d\zeta_1d\zeta_2 \\
   & =
  \int_{\R\times\R_+} \int_{\R\times\R_+}
  \delta(\xi_1- \xi_2) \delta(\vartheta_1- \vartheta_2)
    \ol{\hatt{f}_1\otimes \hatt{f}_2}(\xi_1,\vartheta_1) \hatt{f}_1\otimes \hatt{f}_2(\xi_2,\vartheta_2)\,
    \frac{d\xi_1d\vartheta_1 d\xi_2 d\vartheta_2}{4|\eta_2-\eta_1||\zeta_2-\zeta_1|} \\
   & =
  \int_{\R\times\R_+}
    |\hatt{f}_1\otimes \hatt{f}_2(\xi_1,\vartheta_1)|^2 \,
    \frac{d\xi_1d\vartheta_1}{4|\eta_2-\eta_1|^2} \\
   & \le
  \frac{1}{2\,\dist(\supp \hatt{f}_1, \supp \hatt{f}_2)}
  \int_{\R\times\R_+}
    |\hatt{f}_1\otimes \hatt{f}_2(\xi_1,\vartheta_1)|^2 \,
    \frac{d\xi_1d\vartheta_1}{2|\eta_2-\eta_1|} \\
    & =
  \frac{1}{2\,\dist(\supp \hatt{f}_1, \supp \hatt{f}_2)}
   \| \hatt{f}_1\|_2^2  \| \hatt{f}_2\|_2^2
 \end{align*}
where, in the last equality, we undid the change of variables
$\xi_1=\eta_1+\eta_2$, $\vartheta_1=\eta_1^2+\eta_2^2$. This proves
\eqref{eq:bilinear-Fourier-space}.
\end{proof}

\begin{remarks}
\itemthm The Fourier space version of the bilinear estimate has a
generalization to arbitrary space dimension. If $f_j\in L^2(\R^d)$
are well separated in Fourier space and $T_t= e^{it\Delta}$, with
$\Delta= \sum_{j=1}^d\partial_{x_j}^2$ the Laplacian in $\R^d$, the
free Schr\"odinger time evolution, then
  \beq\label{eq:bilinear-Fourier-space-general-dim}
  \| T_tf_1 T_tf_2\|_{L^2(\R\times\R^d, dtdx)} \le
  \frac{C}{\sqrt{\dist (\supp \hatt{f_1}, \supp \hatt{f_2})}}
  \| f_1\|_{L^2(\R^d)} \| f_2\|_{L^2(\R^d)} .
  \eeq
for some constant depending on $d$, see, for example,
\cite{KPV91,CKSTT01}.\\[0.2em]
\itemthm Similarly, a suitable version of the duality Lemma is valid
in all space dimensions. To formulate this, let $\hatt{f}$ be the
$d$-dimensional Fourier transform of $f$, given by
  \bdm
   \hatt{f}(\eta)
   =
   \frac{1}{(2\pi)^{d/2}} \int_{\R^d} e^{-ix\eta} f(x)\, dx
  \edm
for $f\in L^1(\R^d)\cap L^2(\R^d)$ and extended to a unitary
operator to all of $L^2(\R^d)$. The inverse Fourier transform is
again denoted by $\check{f}$,
 \bdm
 \check{ f} (x)= \f{1}{(2\pi)^{d/2}}\int_{\R^d} e^{ix\eta}f(\eta)d\eta \,\,.
 \edm
Then
 \beq\label{eq:duality-bilinear-general-dim}
\| T_tf_1 T_t f_2\|_{L^2(\R\times \R^d, |t|^{d-2} dt dx)}
  =
  2^{1-d/2} \| T_t\check{f_1} T_t \check{f_2}\|_{L^2(\R\times \R^d, dt
  dx)}.
 \eeq
This follows from a similar calculation as in the proof of Lemma
\ref{lem:duality-bilinear} using now the representations
 \begin{align}\label{eq:freetimeevolution-general-dim}
   T_t f(x)
  =
   \frac{1}{(4\pi it)^{d/2}} \int_{\R^d} e^{i\frac{|x-y|^2}{4t}} f(y)\, dy
  =
   \frac{1}{(2\pi)^{d/2}} \int_{\R^d} e^{ix\eta} e^{-it\eta^2} \hatt{f}(\eta)\, d\eta
 \end{align}
for the free Schr\"odinger evolution in $\R^d$.\\[0.2em]
\itemthm The bilinear estimate for initial conditions which are
separated  in $x$-space is our main tool to get decay estimates on
the dispersion managed soliton in $x$-space. By the two remarks
above, the proof of Lemma \ref{lem:bilinear-real} immediately
generalizes to all space dimensions giving the following bilinear
real space estimate: If $f_j\in L^2(\R^d)$ have separated supports
and $T_t= e^{it\Delta}$, with $\Delta= \sum_{j=1}^d\partial_{x_j}^2$
the Laplacian in $\R^d$, the free Schr\"odinger time evolution, then
 \beq\label{eq:bilinear-real-space-general-dim}
 \| T_tf_1 T_t f_2\|_{L^2(\R\times \R^d, |t|^{d-2} dt dx)}
  \le
  \frac{C}{\sqrt{\dist (\supp f_1 , \supp f_2)}}
  \| f_1\|_{L^2(\R^d)} \| f_2\|_{L^2(\R^d)}
 \eeq
for some constant $C$.\\[0.2em]
\itemthm The duality under Fourier transform in Lemma
\ref{lem:duality-bilinear} was first noticed in the context of the
Strichartz estimate in \cite{HZ06} for the Strichartz norm in
dimension one and two. In fact, it holds in general for suitable
mixed space-time norms
 \beq\label{eq:mixed-space-time}
  \|u\|_{L^r_tL^p_x}=\Big(\int_\R\Big(\int_{\R^d}|u(t,x)|^p\,
  dx\Big)^{r/p}dt\Big)^{1/r} .
 \eeq
If $u(t,x)= T_tf(x)$ is the solution of the free Schr\"odinger
equation in $\R^d$ then, using
\eqref{eq:freetimeevolution-general-dim}, a similar change of
variables calculation as in the proof of Lemma
\ref{lem:duality-bilinear} yields the symmetry
 \beq
  \|T_tf\|_{L^r_tL^p_x} = \|T_t \check{f}\|_{L^r_tL^p_x}
  \quad \text{ for }\frac{2}{r}= \frac{d}{2} - \frac{d}{p} \,\,.
 \eeq
The observation made here, that this type of invariance immediately
transforms Fourier-space bilinear estimates into corresponding
$x$-space bilinear bounds seems to be new.\\[0.2em]
\itemthm In a forthcoming paper, \cite{HuLee08}, we use the Fourier and $x$-space
bilinear Strichartz estimates to give a simple proof of existence of
minimizers of the minimization problem \eqref{eq:minimization-critical} which
avoids the use of Lion's concentration compactness principle.
\end{remarks}

For the application we have in mind, we need to have similar
estimates for the functional $\calQ$.

\begin{cor}[Multi-linear estimates]\label{cor:4linear} Let $f_j\in
L^2(\R)$ for $j=1,2,3,4$.\\[0.2em]
\itemthm If there exists a pair $i \not=j$ such that $\dist (\supp
f_i, \supp f_j)>0$, then
 \begin{equation}\label{eq:multi-linear-real-space}
 | \calQ(f_1, f_2, f_3, f_4) |
 \le
 \f{1}{2^{1/4} \, 3^{3/8}\,\sqrt{\dist (\supp f_i, \supp f_j)}}
 \|f_1\| \|f_2\|\|f_3\| \|f_4\|.
 \end{equation}
\itemthm If there exists a pair $i \not=j$ such that $\dist (\supp
\hatt{f_i}, \supp \hatt{f_j})>0$, then
 \begin{equation}\label{eq:multi-linear-Fourier-space}
 | \calQ(f_1, f_2, f_3, f_4) |
 \le \f{1}{2^{3/4} 3^{1/8}\sqrt{\dist (\supp \hatt{f_i},
 \supp \hatt{f_j})}} \|f_1\| \|f_2\|\|f_3\| \|f_4\|.
 \end{equation}
\end{cor}
\begin{proof}
Since
$|\calQ(f_1,f_2,f_3,f_4)|\le \int_0^1 \int_{\R} |T_tf_1
T_tf_2T_tf_3T_tf_4|\, dt dx$ we can, without loss of generality,
assume $i=1$ and $j=2$. First we prove
\eqref{eq:multi-linear-real-space}. Using the Cauchy-Schwarz
inequality,
 \begin{align}
 \calQ(f_1,f_2,f_3,f_4) & \le \int_0^1 \int_{\R} |T_tf_1T_tf_2T_tf_3T_tf_4|\, dxdt \nonumber \\
  & \le
   \big( \int_0^1 \int_{\R} \f{|T_tf_1T_tf_2|^2}{t}\, dxdt  \big)^{1/2}
   \big( \int_0^1 \int_{\R} t |T_tf_3T_tf_4|^2\, dxdt  \big)^{1/2} .
   \label{eq:first-factor}
 \end{align}
The first factor is bounded by \eqref{eq:bilinear-real-space}. The
second factor equals $(\calR(f_3,f_3,f_4,f_4))^{1/2}$, which is
bounded by \eqref{eq:R}. This shows \eqref{eq:multi-linear-real-space}.
The proof of \eqref{eq:multi-linear-Fourier-space} is analogous, using
\eqref{eq:bilinear-Fourier-space} and \eqref{eq:Q}.
\end{proof}

\begin{remark} We always have the bound $|\calQ(f_1,f_2,f_3,f_4)|\le P_1 \|f_1\|
\|f_2\|\|f_3\| \|f_4\|$ by Lemma \ref{lem:0}. So the bounds
\eqref{eq:multi-linear-real-space} and
\eqref{eq:multi-linear-Fourier-space} can be improved for small separation
of the supports. Chasing the constants, one sees
 \begin{equation}\label{eq:multi-linear-real-space-refined}
 | \calQ(f_1, f_2, f_3, f_4) |
 \le
 P_1 \min\big(1,\f{1.33}{\sqrt{\dist (\supp f_i, \supp
 f_j)}}\big)
 \|f_1\| \|f_2\|\|f_3\| \|f_4\|
 \end{equation}
and
 \begin{equation}\label{eq:multi-linear-Fourier-space-refined}
 | \calQ(f_1, f_2, f_3, f_4) | \le P_1\min\big(1,\f{1.1}{\sqrt{\dist (\supp \hatt{f_i},
 \supp \hatt{f_j})}}\big)\|f_1\| \|f_2\|\|f_3\| \|f_4\|
 \end{equation}
but for our purposes precise estimates for the constants are not needed since
they only indirectly affect the bound on the decay rate, see the proof of
Corollary \ref{cor:arbitrary-polyn-decay}.
\end{remark}
The next result is the second main ingredient for our bounds on
dispersion managed solitons. It shows that although the functional
$\calQ(f_1,f_2,f_3,f_4)$ is highly non-local, it retains at least
some locality both in Fourier and $x$-space.

\begin{lemma}[Quasi-locality of $\calQ$] \label{lem:quasi-locality} Let $s>0$
and $i=1,2,3$ or $4$.   \\[0.2em]
\itemthm If $\supp f_i \subset \{|x| > 3s\}$ and $\supp f_j \subset\{|x|\le s\}$
for all $j \neq i$, then
 \bdm
  \calQ(f_1, f_2, f_3, f_4)=0.
 \edm\\[-0.3em]
\itemthm If $\supp \hat f_i \subset \{|k|> 3s\}$ and $\supp \hat f_j \subset
\{|k|\le s\}$ for all $j \neq i$, then
 \bdm
  \calQ(f_1, f_2,f_3,f_4)=0.
 \edm
\end{lemma}

\bpf
We give the proof for $i=1$, the other cases are similar. For part
(i) of the lemma, we express $\calQ(f_1, f_2, f_3, f_4)$ using
\eqref{eq:freetimeevolution1} similar to the proof of the Duality
Lemma \ref{lem:duality-bilinear}.
 \begin{align}\label{eq:quasi-local-real}
 & \calQ(f_1, f_2, f_3, f_4) \nonumber \\
 & = \int_0^1
 \f{d t}{(4\pi t)^2} \int _{\R} d x \int _{\R^4}
 e^{\f{ix(y_1 -y_2+y_3 -y_4)}{2t}}
 e^{\f{-i(y_1^2-y_2^2+y_3^2-y_4^2)}{4t}}
 \ol{f_1(y_1)}f_2(y_2)\ol{f_3(y_3)}f_4(y_4)
 d y \nonumber \\
 & = \f{1}{8 \pi ^2}\int_0^1
 \f{d t}{t} \int _{\R} d z \int _{\R^4}
 e^{i(y_1 -y_2+y_3 -y_4)z}
 e^{\f{-i(y_1^2-y_2^2+y_3^2-y_4^2)}{4t}}
 \ol{f_1(y_1)}f_2(y_2)\ol{f_3(y_3)}f_4(y_4)
 d y \nonumber \\
 & = \f{1}{4 \pi}\int_0^1
 \f{d t}{t} \int _{\R^4}
 \delta(y_1-y_2+y_3-y_4)
 e^{\f{-i(y_1^2-y_2^2+y_3^2-y_4^2)}{4t}}
 \ol{f_1(y_1)}f_2(y_2)\ol{f_3(y_3)}f_4(y_4)
 d y
 \end{align}
where we made the change of variables with $x=2tz$. The
$\delta$-functions restrict the integration to the subspace
$y_1=y_2-y_3+y_4$. Because of our assumption on the supports of
$f_j$, the product of the $f_j$, and hence the integrand, vanishes
for any $(y_1, y_2, y_3, y_4)$ with $y_1=y_2-y_3+y_4$. This proves
part (i).

Analogously, for part (ii), we use the representation
\eqref{eq:freetimeevolution1} to see
 \begin{align*}
 & \calQ(f_1, f_2, f_3, f_4) \\
 & = \f{1}{(2\pi)^2}\int_0^1
 dt \int _{\R} d x \int _{\R^4}
 e^{-ix(\eta_1 -\eta_2+\eta_3 -\eta_4)}
 e^{it(\eta_1^2-\eta_2^2+\eta_3^2-\eta_4^2)}
 \ol{\hatt{f}_1(\eta_1)}\hatt{ f}_2(\eta_2)\ol{\hatt{f}_3(\eta_3)}\hatt{ f}_4(\eta_4)
 d \eta \\
 & = \f{1}{2 \pi}\int_0^1
 d t  \int _{\R^4}
 \delta(\eta_1-\eta_2+\eta_3-\eta_4)
 e^{it(\eta_1^2-\eta_2^2-\eta_3^2-\eta_4^2)}
 \ol{\hatt{f}_1(\eta_1)}\hatt{f}_2(\eta_2)\ol{\hatt{ f}_3(\eta_3)}\hatt{ f}_4(\eta_4)
 d \eta \\
 & = 0
 \end{align*}
under the condition of the support of $\hatt{f}_j$, $j=1,2,3,4$.

\epf
\begin{remarks}
\itemthm As the above proof shows, $\calQ(f_1,f_2,f_3,f_4)=0$ if
either $0\not\in \supp(f_1)-\supp(f_2)+\supp(f_3)-\supp(f_4)$ or
$0\not\in
\supp(\hatt{f}_1)-\supp(\hatt{f}_2)+\supp(\hatt{f}_3)-\supp(\hatt{f}_4)$.\\[0.2em]
\itemthm That the functional $\calQ$ is quasi-local in Fourier space is
not necessarily a surprise. In Fourier space the space integral of
the product of the time evolved wave packets  $T_tf_j$ amounts to a
convolution of the respective Fourier transforms. The additional
$\delta$-function in the variables $\eta_1-\eta_2+\eta_3-\eta_4$
expressed momentum conservation, since $\calQ$ is invariant under
translations. That the same result holds for wave packets
corresponding to initial conditions which are separated in real
space is more surprising, since the free Schr\"odinger equation is
dispersive and the wave packets $T_tf_j$ have a lot of overlap for
$t\not=0$, even if they are well separated for $t=0$. \\[0.2em]
\itemthm There is a related duality result for $\calQ$ similar to the
duality Lemma \ref{lem:duality-bilinear} for the bilinear norms, which
explains a bit the quasi-locality of $\calQ$ in real space. For
this it is natural to consider a more general class of functionals
given by
 \bdm
  \calQ_{\psi}(f_1,f_2,f_3,f_4) =
  \int_{\R} \int_{\R} \ol{T_tf_1(x)} T_tf_2(x) \ol{T_tf_3(x)} T_tf_4(x) \,\psi(t) dx dt
 \edm
for a suitable cuff-off function $\psi$. Similar to the proof of
Lemma \ref{lem:0}, it is easy to see that $\calQ$ is bounded on
$L^2(\R)$. Expressing $T_tf_j$ via \eqref{eq:freetimeevolution1} one
has
 \begin{multline*}
  \calQ_\psi(f_1,f_2,f_3,f_4) \\
  = \frac{1}{(4\pi)^2}\int_{\R} \int_{\R}
     \int_{\R^4} e^{ix(y_1-y_2+y_3-y_4)/(2t)} e^{-i(y_1^2-y_2^2+y_3^2-y_4^2)/(4t)} \\
     \phantom{abcdefghijklmnop}\ol{f_1(y_1)} f_2(y_2) \ol{f_3(y_3)} f_4(y_4)\, dy
   \,\f{\psi(t)}{|t|^{2}} dx dt \\
  =
   \frac{1}{(2\pi)^2}\int_{\R} \int_{\R}
     \int_{\R^4} e^{iz(y_1-y_2+y_3-y_4)} e^{-i\tau(y_1^2-y_2^2+y_3^2-y_4^2)} \\
     \ol{f_1(y_1)} f_2(y_2) \ol{f_3(y_3)} f_4(y_4) \, dy
   \,\f{\psi(1/(4\tau))}{2|\tau|} dz d\tau
 \end{multline*}
where we first changed variables $x=2tz$, $dx= 2|t|dz$ and then
$\tau=1/(4t)$, $\tfrac{d\tau}{|\tau|} = \tfrac{d t}{|t|}$. Hence
with $\wti{\psi}(\tau)=\psi(1/(4\tau))/(2|\tau|)$ and recalling
$\eqref{eq:freetimeevolution2}$,
 \beq\label{eq:duality-multi-linear}
  \calQ_\psi(f_1,f_2,f_3,f_4) = \calQ_{\wti{\psi}}(\check{f}_1,\check{f}_2,\check{f}_3,\check{f}_4)
 \eeq
where $\check{f}_j$ is the inverse Fourier transform of $f_j$. In
particular, any result on $\calQ_{\wti{\psi}}$ under conditions on
the Fourier transforms of the involved functions implies the same
result for $\calQ_{\psi}$ under \emph{exactly} the same conditions
on the original functions $f_j$. For example, quasi-locality of $\calQ_{\wti{\psi}}$
in Fourier space is equivalent to quasi-locality of $\calQ_{\psi}$ in real space.
\end{remarks}

\section{Proof of the main result} \label{sec:proof-main}
Let $f\in L^2(\R)$ and recall the tail distributions $\alpha(s)=
(\int_{|x|>s}|f(x)|^2\, dx)^{1/2}$ and $\beta(s)
=(\int_{|k|>s}|\hatt{f}(k)|^2\, dk)^{1/2}$.
Our main tool for proving the decay estimates for dispersion managed
solitons is the following self-consistency bound on the tail
distribution. For two functions $g$ and $h$ we write $g\lesssim h$
if there exists a constant $C>0$ such that $g\le Ch$.

\begin{lemma}[Self-consistency estimate]\label{lem:tail-estimate}
Let $\omega>0$ and $f\in L^2(\R)$ be a weak solution of
$\omega f=Q(f,f,f)$. Denote by $\alpha$, respectively $\beta$, the
tail distributions of $f$, respectively its Fourier transform. Then
for all $s>0$
 \beq \label{eq:selfconsistency1}
 \alpha(3s) \lesssim
 (\alpha(s))^3 + \frac{\alpha(0)^2 \alpha(s)}{\sqrt{s}}
 \eeq
and
 \beq \label{eq:selfconsistency2}
 \beta(3s) \lesssim
 (\beta(s))^3 + \frac{\beta(0)^2 \beta(s)}{\sqrt{s}} .
 \eeq
The implicit constant in the above estimates is bounded by
$CP_1/\omega$ for some absolute constant $C$.
\end{lemma}

\begin{remarks}
\itemthm One can improve on this a little bit by replacing $\sqrt{s}$
with $\max(\sqrt{s},1)$ and one of the factors $\alpha(0)$,
respectively $\beta(0)$, by $\alpha(0)-\alpha(s)$,
respectively $\beta(0)- \beta(s)$
in the above bounds. As the proof of Corollary \ref{cor:arbitrary-polyn-decay} shows,
however, the precise value of the constant in the self-consistency bounds is not relevant
for the decay estimates.\\[0.2em]
\itemthm With \eqref{eq:omega-ground-state} for the ground--state soliton and using
\eqref{eq:multi-linear-real-space-refined} and \eqref{eq:multi-linear-Fourier-space-refined}
to chase the constants in the proof of Lemma \ref{lem:tail-estimate}  one sees
that the rather explicit bounds
 \beq \label{eq:selfconsistency1-constants}
  \ol{\alpha}(3s) \le
  (\ol{\alpha}(s))^3 + 3\min(1,\frac{1}{\sqrt{s}} ) (1-\ol{\alpha}(s))\ol{\alpha}(s)
 \eeq
and
 \beq \label{eq:selfconsistency2-constants}
 \ol{\beta}(3s) \le
  (\ol{\beta}(s))^3 + 3\min(1,\frac{0.78}{\sqrt{s}} ) (1-\ol{\beta}(s))\ol{\beta}(s)
 \eeq
for the normalized tail distributions
$\ol{\alpha}(s)=\alpha(s)/\alpha(0)$, respectively
$\ol{\beta}(s)=\beta(s)/\beta(0)$, of the ground--state soliton hold.  In the
limit $s\to 0$, these bounds cannot be improved. \\[0.2em]
\itemthm The self--consistency bounds for $\alpha$ and $\beta$
provided by Lemma \ref{lem:tail-estimate} are instrumental for our
proof that $\alpha$ and $\beta$ decay faster than any
polynomial at infinity. The key property for this, as expressed by the the bounds
\eqref{eq:selfconsistency1} and \eqref{eq:selfconsistency2}, is the somewhat
surprising fact that, despite the dispersion management equation being a highly non-local
equation, the values of any weak solution of $f=Q(f,f,f)$ on the set $\{|x|>3s\}$ can be
controlled solely by the values of $f$ on the slightly enlarged set $\{|x|>s\}$.
This important property is due to the quasi-locality of $\calQ$, as expressed in
Lemma \ref{lem:quasi-locality}.\\[0.2em]
\itemthm Although the self-consistency bounds \eqref{eq:selfconsistency1} and \eqref{eq:selfconsistency2}
are not strong enough to yield exponential decay of $\alpha$ and
$\beta$, they are not too far from the truth: A bound of the
form
 \beq\label{eq:best-selfconsistency}
  \alpha(3s)\lesssim (\alpha(s))^3 \quad \text{and } \beta(3s)\lesssim
  (\beta(s))^3,
 \eeq
 i.e., dropping the second term, together with some decay of $\alpha$ can be
 bootstrapped to yield exponential decay of both $\alpha$ and $\beta$,
 see Remark \ref{rem:exponential}.\\[0.2em]
\end{remarks}
\bpf[Proof of the self-consistency bounds] First we prove \eqref{eq:selfconsistency1}. Fix $s>0$. Recall that
$f$ is a weak solution of $f=Q(f,f,f)$ if and only if $\la g,f \ra=
\calQ(g,f,f,f)$ for all $g\in L^2(\R)$. Since the left hand side of
\eqref{eq:selfconsistency1} is
 \beq \label{eq:innerproduct}
 \alpha(3s)= \sup_{\substack{\supp(g)\subset (-\infty,-3s)\cup(3s,\infty) \\
                \|g \|=1}}|\la g,f\ra| ,
 \eeq
it remains to estimate $\calQ(g,f,f,f)$ uniformly in $g \in L^2(\R)$
with $\supp g \subset (-\infty,-3s)\cup(3s,\infty)$ and $\|g\|_2=1$.
Let $I_s=[-s,s]$. We split $f$ into its low and high space
parts, according to $I_s$: $f_<= f_{<,s}= f \chi_{I_s}$ and $f_>=
f_{>,s}= f (1-\chi_{I_s})$, where $\chi_{I_s}$ is the characteristic
function of the interval $I_s$, and use the multi-linearity of
$\calQ$ to rewrite
 \begin{align}\label{eq:estimate}
 \omega \la g,f \ra  &=  \calQ(g,f,f,f)
   = \calQ(g, f_{<}, f_{<}, f_{<})
 + \calQ(g, f_{>}, f_{>}, f_{>}) \no \\
 & \phantom{\,\, =  \calQ(g,f,f,f)=} + \calQ(g,f,f_{<},f_{>}) + \calQ(g,f_{>},f,f_{<})
 +\calQ(g,f_{<},f_{>},f)\no\\
 & = \calQ(g, f_{>}, f_{>}, f_{>})
  + \calQ(g,f,f_{<},f_{>}) + \calQ(g,f_{>},f,f_{<})
  +\calQ(g,f_{<},f_{>},f)
 \end{align}
where the last equality follows from the quasi-locality, $\calQ(g,
f_{<}, f_{<}, f_{<})=0$ from Lemma \ref{lem:quasi-locality}, since
the supports of $g$ and $f_<$ do not match by the definition of
$f_<$.

Using Lemma \ref{lem:0}, the first term on the right hand side of \eqref{eq:estimate}
is bounded by
 \bdm
  |\calQ(g,f_>,f_>,f_>)| \lesssim \|g\|_{2} \|f_>\|_2^3 = (\alpha(s))^3.
 \edm
It remains to bound the last three terms of \eqref{eq:estimate}.
Since, by assumption, the supports of $g$ and $f_<$ are separated by
at least $2s$, we can use  Corollary \ref{cor:4linear} to see
 \bdm
 | \calQ(g,f,f_{<},f_{>}) |
 \lesssim \f{1}{\sqrt{s}} \|f\| \| f_{<} \| \| f_{>} \|
 \le \f{1}{\sqrt{s}} \|f\|^2 \| f_{>} \| = \f{1}{\sqrt{s}}\alpha(0)^2\alpha(s) .
 \edm
The bounds for the other two terms are the same.

To prove the bound \eqref{eq:selfconsistency2}, one notes for any $\hat
g \in L^2(\R)$, $\la \hat g, \hat f\ra = \la g, f \ra =
\calQ(g,f,f,f)$. Since
 \bdm
 \beta(3s)= \sup_{\substack{\supp(\hatt{ g})\subset (-\infty,-3s)\cup(3s,\infty) \\
                        \|g \|=1}}|\la \hatt{ g}, \hatt{ f}\ra|.
 \edm
a proof similar to the above one, splitting $f$ into its low and
high frequency parts,  gives the bound \eqref{eq:selfconsistency2} for
$\beta$. \epf
A-priori we only know that $\alpha$ and $\beta$ decay to
zero as $s\to\infty$ for an arbitrary $f\in L^2(\R)$. The self
consistency bounds of Lemma \eqref{lem:tail-estimate} allow us to
bootstrap this and get some explicit super-polynomial decay. To see
how this might work, assume for the moment that $h:\R_+\to \R_+$
obeys the bound
 \bdm
  h(s)\lesssim (h(s))^3
 \edm
for all $s\in\R_+$. Then, of course, for all $s\in\R_+$ either
$h(s)=0$ or $1\lesssim h(s)$. So if  in addition one knows
that $h$ decays to zero at infinity, it must already have
compact support. The following makes this intuition precise.
\begin{cor}[$=$ strengthening of Proposition
\ref{prop:decay}]\label{cor:arbitrary-polyn-decay} Let $\omega >0$ and
$f\in L^2(\R)$ a weak solution of $\omega f=Q(f,f,f)$. Then there exist $s_0$,
respectively $\hatt{s}_0$, such that
 \bdm
 \begin{split}
 \alpha(s) &\le \alpha(s_0) 3^{1/4} 3^{-(\log_3(\f{s}{3s_0}))^2/4} , \\
 \beta(s) &\le \beta(\hatt{s}_0) 3^{1/4} 3^{-(\log_3(\f{s}{3\hatt{s}_0}))^2/4},
 \end{split}
 \edm
for all $s\ge s_0$, respectively $s\ge \hatt{s}_0$.
\end{cor}
\begin{remarks}
\itemthm The above bounds are only effective when $s\ge 9s_0$, respectively $s\ge 9\hatt{s}_0$.
Since $\alpha$ and $\beta$ are monotone decreasing, they are bounded
by $\alpha(0)= \beta(0) = \|f\| $ for small $s$.\\[0.2em]
\itemthm Using Remark \ref{rem:f-decay}, we get the same point-wise decay estimate for $f$.\\[0.2em]
\itemthm The value of $s_0$, which is the only quantity in the decay estimate affected
by the value of the constant in the self-consistency bound from Lemma
\ref{lem:tail-estimate}, is determined in \eqref{eq:bootstrap-start} below.
\end{remarks}
\bpf
We prove only the first bound, the proof for  the second is
identical. By \eqref{eq:selfconsistency1}, we know that
 \beq \label{eq:bootstrap0}
    \alpha(3s) \le C\Big(\alpha(s)^2 + \frac{\alpha(0)^2}{\sqrt{s}}\Big) \alpha(s)
 \eeq
for some constant $C$. Since $f\in L^2(\R)$, $\alpha$ is
monotonically decreasing with $\alpha(\infty)=
\lim_{s\to\infty}\alpha(s)=0$. Thus there exists $s_0<\infty$ such that
 \beq\label{eq:bootstrap-start}
   C\Big(\alpha(s_0)^2 + \frac{\alpha(0)^2}{\sqrt{s_0}}\Big) \le 3^{-1/4} .
 \eeq
The monotonicity of $\alpha$ together with \eqref{eq:bootstrap-start} and \eqref{eq:bootstrap0}
yield the a-priori bound
 \bdm
  \alpha(3s)
  \le
  3^{-1/4} \alpha(s) \quad\text{ for all } s\ge s_0 .
 \edm
Putting $\gamma(t):= \log_3(\alpha(3^t))$ and $t_0=\log_3(s_0)$, we see that
 \beq\label{eq:bootstrap2}
  \gamma(t+1)\le \gamma(t) - \f{1}{4} \quad \text{ for all }  t\ge t_0 .
 \eeq
With $\wti{\gamma}_1(t)= \gamma(t)+t/4$ this is equivalent to
 \bdm
  \wti{\gamma}_1(t+1)- \wti{\gamma}_1(t)= \gamma(t+1)-\gamma(t) + \f{1}{4}\le 0 \quad \text{ for all }  t\ge t_0,
 \edm
which shows that $\wti{\gamma}_1$ is sub-periodic for $t\ge t_0$.
In particular,
 \bdm
   \wti{\gamma}_1(t)\le \sup_{t'\in [t_0, t_0+1)} \wti{\gamma}_1(t')
   \le \gamma(t_0) + \f{t_0+1}{4} = \log_3\big(\alpha(s_0)(3s_0)^{1/4}\big)
 \edm
for all $t\ge t_0$ since $\alpha(s)$ and hence also $\gamma(t)$ is decreasing. In turn, this
yields $\gamma(t)\le \log_3\big(\alpha(s_0)(3s_0)^{1/4}\big) -\f{t}{4}$, or, equivalently,
 \beq\label{eq:bootstrap5}
  \alpha(s)\le  \alpha(s_0) \Big(\frac{ 3s_0}{s}\Big)^{1/4} \quad \text{ for all }  s\ge s_0  .
 \eeq
Now we bootstrap this once. Plugging \eqref{eq:bootstrap5} back into \eqref{eq:bootstrap0} and using
\eqref{eq:bootstrap-start} one gets
 \bdm
 \begin{split}
   \alpha(3s)
  &\le
   C\Big( \alpha(s_0)^2  + \frac{\alpha(0)^2}{\sqrt{3s_0}}\Big)
   \Big(\frac{3s_0}{s}\Big)^{1/2} \alpha(s) \\
  &\le
   3^{-1/4} \Big(\frac{3s_0}{s}\Big)^{1/2} \alpha(s)
 \end{split}
 \edm
 for all $s\ge s_0$.
Hence \eqref{eq:bootstrap2} is improved to
 \beq\label{eq:bootstrap7}
  \gamma(t+1)\le \gamma(t)-\frac{1}{4} + \frac{t_0+1- t}{2} \quad \text{ for all } t\ge t_0 .
 \eeq
With $\wti{\gamma}_2(t):=\gamma(t) + (t-t_0-1)^2/4 $ the bound \eqref{eq:bootstrap7} is equivalent to
 \bdm
   \wti{\gamma}_2(t+1)- \wti{\gamma}_2(t)\le 0
   \quad \text{ for all } t\ge t_0 .
 \edm
Hence, for all $t\ge t_0$,
 \bdm
   \wti{\gamma}_2(t)  \le \sup_{t\in [t_0,t_0+1]}\wti{\gamma}_2(s)
   \le
    \gamma(t_0) + \frac{1}{4}.
 \edm
Equivalently,
 \bdm
  \gamma(t)\le \gamma(t_0) +\frac{1}{4} - \frac{(t-t_0-1)^2}{4}  \quad\text{ for all } t\ge t_0,
 \edm
which yields the claimed inequality for $\alpha(s)$. Given
\eqref{eq:selfconsistency2}, the same proof applies to the tail
distribution of $\hatt{f}$. \epf

\begin{remarks}\label{rem:decay}
 \itemthm
 There are non exponentially decaying functions which obey the
 self-consistency bounds of Lemma \ref{lem:tail-estimate}. For example,
 \bdm
   g(s)= 3^{-(\log_3(\f{s}{3s_0}))_+^2},
 \edm
 with $(r)_+= \max(0,r)$, obeys the bound
 \bdm
   g(3s)\le \sqrt{\frac{3s_0}{s}}\, g(s)
 \edm
 for all $s>0$. Thus the bounds given in Lemma
 \ref{lem:tail-estimate} are not strong enough to yield the conjectured
 exponential decay for the dispersion managed soliton.\\[0.2em]
 \itemthm \label{rem:exponential} The self-consistency bounds given
 by Lemma \ref{lem:tail-estimate} are not too far from the truth.
 A bound of the form
 \beq\label{eq:best-selfconsistency2}
   \alpha(3s)\lesssim \alpha(s)^3
 \eeq
 is not only consistent with exponential decay of $\alpha$, but, together with
 the a-priori decay $\lim_{s\to\infty}\alpha(s)=0$, implies exponential decay of $\alpha$.
 To see this, let us assume first
 \bdm
   \alpha(3s)\le \alpha(s)^3
 \edm
 for all $s\ge 0$. With $\gamma(t)=\log_3\alpha(3^t)$, this is equivalent to
 \bdm
   \gamma(t+1)\le 3\gamma(t)
 \edm
 for all $t$ and  iterating this bound yields
 \beq\label{eq:super3}
   \gamma(t) \le 3^n\gamma(t-n)\quad \text{for all } t \text{ and all } n\in\N_0.
 \eeq
 Since $\gamma(t)\to-\infty$, as $t\to\infty$, we can
 choose $t_0$ such that
 \bdm
   3\mu := -\gamma(t_0) >0 .
 \edm
 With this choice \eqref{eq:super3} implies
 \bdm
   \gamma(t_0+n)\le -\mu 3^{n+1}
 \edm
 for all $n$. Since $\alpha$ and hence $\gamma$ is decreasing, this gives
 \bdm
   \gamma(t) \le -\mu 3^{n+1} \quad \text{for all } t\in [t_0+n, t_0+n+1]
 \edm
 or, equivalently,
 \bdm
   \alpha(s)\le 3^{-\mu 3^{n+1}} \quad \text{for all } s\in [s_0 3^n, s_03^{n+1}]
 \edm
 where $s_0 = 3^{t_0}$. Thus
 \bdm
   \alpha(s)\le 3^{-\mu s/s_0} \quad \text{for all } s\ge s_0 .
 \edm
 If $\alpha(3s)\le C \alpha(s)^3$ for all $s\ge 0$ with $C>0$, then,
 with $\wti{\alpha}(s)= \sqrt{C}\alpha(s)$,
 \bdm
   \wti{\alpha}(3s)\le \wti{\alpha}(s)^3 \quad\text{for all } s\ge 0.
 \edm
 By the above argument $\wti{\alpha}$, hence also $\alpha$, decays exponentially if
 a bound of the form \eqref{eq:best-selfconsistency2} holds.
\end{remarks}

For the last two results, which finish the proof of
Proposition \ref{prop:smooth} and Theorem \ref{thm:main}, it
is convenient to introduce one more notation: for $x\in\R$ let
$\la x\ra =\sqrt{1+x^2}$.

\begin{cor}[$=$ Proposition \ref{prop:smooth}]
If $f\in L^2(\R)$ is a weak solution of $\omega f= Q(f,f,f)$
with $\omega>0$, then the functions $x\mapsto\la x\ra^n f(x)$
and $k\mapsto \la k\ra^m\hatt{f}(k)$ are both square integrable
for all $ m,n\in\N$. In particular, both $f$ and its Fourier
transform are $C^\infty$ functions with all their derivatives,
of arbitrary order, square integrable functions.
\end{cor}
\bpf From Corollary \ref{cor:arbitrary-polyn-decay} we know
$\beta(s)^2= \int_{|k|>s}|\hatt{f}(k)|^2\, dk$ decays faster
than any polynomial. Thus
 \bdm
 \begin{split}
  \int_\R \la k\ra^{m} |\hat{f}(k)|^2\, dk
 &=
  - \int_0^\infty \la s\ra^m d(\beta(s)^2) \\
 &=
  \int_0^\infty m\la s\ra^{m-2}s (\beta(s))^2\, ds +
  \beta(0)^2 <\infty,
 \end{split}
 \edm
 the integration by parts is justified due to the super-polynomial
 decay of $\beta$. The argument for $x\mapsto \la x\ra^n
 f(x)$ is identical. Thus both $f$ and $\hatt{f}$ are in all the Sobolev
 spaces $H^s(\R)$ for arbitrary $s>0$ and the smoothness of $f$ and $\hatt{f}$
 follows from the Sobolev embedding theorem.
\epf

These two corollaries together with the following lemma finish the
proof of our main Theorem \ref{thm:main}.

\begin{lemma}\label{lem:Schwartz}
A function $f:\R\to \C$ is a Schwartz function if and only if
$x\mapsto \la x \ra^n f(x)$ is square integrable for all $n\in\N$
and all weak derivatives of $f$ are square integrable.
\end{lemma}
\bpf
Let $D=-i\partial_x$.  Lemma 1 on page 141 in \cite{ReSi1} tells us
that $f$ is a Schwartz function, that is,
 \bdm
  \| f\|_{n,m,\infty} = \sup_{x\in\R}|\la x \ra^n D^m f(x)| <\infty
 \edm
for all $n,m\in\N_0$, if and only if
 \bdm
  \| f\|_{n,m,2} = \Big(\int | \la x \ra^n D^m f(x)|^2 \, dx \Big)^{1/2}
  < \infty
 \edm
for all $n,m\in\N_0$. In particular, for any Schwartz function $f$,
$\| f\|_{n,0,2}$ and $\| f\|_{0,m,2}$ are finite for all $m,n\in\N_0$,
that is, the functions $x\mapsto \la x \ra^n f(x)$ and $x\mapsto D^m f(x)$
are both square integrable for any $n,m\in\N_0$.

To prove the converse, it is clearly enough to show that for all
$m,n\in\N$ finiteness of
$\| f\|_{2n,0,2}$ and $\| f\|_{0,m+j,2}$ for $j=0,1,\ldots,m$
imply $\|f\|_{n,m,2}<\infty$. For any $\eps >0$ define
$\la x \ra _{\veps}=\la x \ra/\la \veps x \ra$.
By Lemma \ref{lem:regularized-japanese} below all derivatives of
$\la x\ra_\veps^{2n}$ are bounded for $0<\veps\le 1$ and all $n\in\N_0$.
In particular, for any $0<\veps\le 1$ and all $ n,m\in \N$ we
also have $\la x\ra_\veps^{2n} g\in H^m(\R)$ as soon as
$g\in H^m(\R)$.

Now let $n,\ m \in \N_0 $ and $f$ such that $\| f\|_{2n,0,2}<\infty$ and
$\| f\|_{0,m+j,2}<\infty$ for $j=0,1,\ldots,m$. In particular,
$f\in H^{2m}(\R)$.
In the following, it is convenient to think of $D^m$ as a self-adjoint
operator with domain $H^m(\R)$.
By the Leibnitz rule for derivatives,
 \begin{align*}
 & \int |\la x \ra ^n_\veps D^m f|^2 dx
   = \la \la x \ra ^n_\veps D^m f, \la x \ra ^n_\veps D^m f \ra
   = \la f, D^m \la x \ra ^{2n}_\veps D^m f \ra \\
 & = \la f, \sum_{j=0}^m
    { m \choose j }
    D^{m-j} \la x \ra ^{2n}_\veps D^{j+m} f \ra
    = \sum_{j=0}^m
    { m \choose j }
 \la  f D^{m-j} \la x \ra ^{2n}_\veps , D^{j+m} f \ra \\
 & \le  \sum_{j=0}^m
    { m \choose j }
    \|f D^{m-j} \la x \ra ^{2n}_\veps \| \|D^{j+m} f\|
    \le
    \sum_{j=0}^m
    { m \choose j }
    \|f \la x \ra ^{(2n-m+j)_+} \| \|D^{j+m} f\| \\
  & \le
    \sum_{j=0}^m
    { m \choose j }
    \|f \|_{2n,0,2} \|f\|_{0,j+m,2}  ,
 \end{align*}
 where the last inequality uses Lemma \ref{lem:regularized-japanese} and
 $\la x \ra ^{(2n-m+j)_+}\le \la x \ra ^{2n}$ for all $j=0,1,\ldots,m$.
Thus, by  monotone convergence,
 \begin{align*}
 \int |\la x \ra ^n D^m f|^2 dx
 =\lim_{\veps\to 0}
 \int |\la x \ra ^n_\veps D^m f|^2 dx
 \le
  \sum_{j=0}^m
    { m \choose j }
    \|f \|_{2n,0,2} \|f\|_{0,j+m,2}
  <\infty
 \end{align*}
by the assumptions on $f$. Hence $\| f\|_{n,m,2}<\infty$.
\epf

To finish the proof of Lemma \ref{lem:Schwartz}, we need the
\begin{lemma}\label{lem:regularized-japanese}
For $\veps\ge 0$ let $\la x\ra_\veps= \frac{\la x\ra}{\la \veps x\ra}$.
Then, with $D=-i\partial_x$,
one has for all $\eta\in\R$ and all $m\in\N_0$
 \beq\label{eq:regularized-japanese}
   |D^m(\la x\ra_\veps^\eta)|
   \lesssim \la x\ra^{(\eta-m)_+}
   = \left\{
                \begin{array}{ccl}
                    \la x\ra^{\eta-m} & \text{for } & m\le \eta \\
                    1 & \text{ for } & m>\eta
                \end{array}
            \right.
 \eeq
 uniformly in $\veps\in[0,1]$ with the implicit constant
 depending only on $\eta$ and $m$.
\end{lemma}
\begin{proof}
A straightforward induction on $m$ shows
that for $j=0,1,\ldots,m$ there are polynomials $p_j= p_{j,m,\eta}$ of degree at most $j$
such that for all $x\in\R$
 \bdm
   D^m(\la x\ra^\eta) = \sum_{j=0}^m p_j(x) \la x\ra^{\eta-m -j} .
 \edm
This immediately implies the bound
 \beq\label{eq:japanese-bound}
   |D^m(\la x\ra^\eta)|\lesssim \la x\ra^{\eta-m}
 \eeq
for all $\eta\in\R$ and all $m\in\N_0$ with a constant depending only on $m$ and $\eta$.
The Leibnitz rule, the triangle inequality, \eqref{eq:japanese-bound}, and the
Binomial formula imply
 \begin{align*}
   |D^m(\la x\ra_\veps^\eta)|
  &= |D^m(\la x\ra^\eta \la \veps x\ra^{-\eta})|
    \le
    \sum_{j=0}^m { m\choose j } |D^{m-j}\la x\ra^\eta| |D^j \la\veps x\ra^{-\eta}| \\
  &\lesssim
    \sum_{j=0}^m { m\choose j } \la x\ra^{\eta-(m-j)} \la\veps x\ra^{-\eta-j} \veps^j
    =
    \frac{\la x\ra^{\eta}}{\la \veps x\ra^{\eta}}
    \sum_{j=0}^m { m\choose j } \la x\ra^{-(m-j)} \la\veps x\ra^{-j} \veps^j \\
  &=
   \frac{\la x\ra^{\eta}}{\la \veps x\ra^{\eta}}
   \Big( \frac{1}{\la x\ra} + \frac{\veps}{\la \veps x\ra}  \Big)^{m}
   \lesssim
   \frac{\la x\ra^{\eta-m}}{\la \veps x\ra^{\eta}}
   =
   \la x\ra_\veps^{\eta-m} \la \veps x\ra^{-m}
   \le \la x\ra_\veps^{\eta-m}
 \end{align*}
 since $\veps \la x\ra\le \la \veps x\ra$ for all $x$ and all $0\le \veps\le 1$.
 This proves \eqref{eq:regularized-japanese} since $1\le \la \veps x\ra \le \la x\ra$
 for all $x$, and $0\le\veps\le 1$.
\end{proof}

\begin{remark}
 Using the multidimensional Binomial and Leibnitz formulas, see Theorem 1.2 in \cite{Raymond},
 the corresponding statement of Lemma \ref{lem:Schwartz} and Lemma
 \ref{lem:regularized-japanese} hold also on $\R^d$  with virtually identical proofs.
\end{remark}

\smallskip\noindent
\textbf{Acknowledgment:} It is  a pleasure to thank Vadim Zharnitsky
for instructive discussions on the dispersion management technique
and introducing us to the problem of decay estimates for dispersion
management solitons.
We would also like to Tony Carbery, Maria and Thomas Hoffmann-Ostenhof,
and Rick Laugesen for discussions.
Young-Ran Lee thanks the School of Mathematics of the University of
Birmingham, UK, for their warm hospitality.

\vspace{5mm}

\def\cprime{$'$}

\end{document}